\documentstyle[aps,prb]{revtex}
\begin{document}
\input epsf.sty
\twocolumn[
\hsize\textwidth\columnwidth\hsize\csname %
@twocolumnfalse\endcsname
%
\widetext


\title{Spin dynamical properties and orbital states of the layered perovskite
La$_{2-2x}$Sr$_{1+2x}$Mn$_{2}$O$_{7}$ $(0.3 \le x < 0.5)$}
\author{K. Hirota}
\address{Department of Physics, Tohoku University, Sendai 980-8578, Japan}
\author{S. Ishihara\cite{S.Ishihara}}
\address{Institute for Materials Research, Tohoku University, Sendai 980-8577, Japan}
\author{H. Fujioka\cite{H.Fujioka}}
\address{Department of Physics, Tohoku University, Sendai 980-8578, Japan}
\author{M. Kubota\cite{M.Kubota} and H. Yoshizawa}
\address{Neutron Scattering Laboratory, Institute for Solid State Physics, University of
Tokyo, Tokai 319-1106, Japan}
\author{Y. Moritomo}
\address{Center for Integrated Research in Science and Engineering, Nagoya University,
Nagoya 464-8601, Japan}
\author{Y. Endoh and S. Maekawa}
\address{CREST, Institute for Materials Research, Tohoku University, Sendai 980-8577,
Japan}
\date{\today}
\maketitle


\begin{abstract}
%
%
Low-temperature spin dynamics of the double-layered perovskite
La$_{2-2x}$Sr$_{1+2x}$Mn$_{2}$O$_{7}$ (LSMO327) was systematically studied in a wide
hole concentration range ($0.3 \le x < 0.5$).
%
%
The spin-wave dispersion, which is almost perfectly two-dimensional, has two branches
due to a coupling between layers within a double-layer.  Each branch exhibits a
characteristic intensity oscillation along the out-of-plane direction.  We found that
the in-plane spin stiffness constant and the gap between the two branches strongly
depend on $x$.   
%
%
By fitting to calculated dispersion relations and cross sections assuming Heisenberg
models,  we have obtained the in-plane $(J_{\parallel})$, {\em intra}-bilayer
$(J_{\perp})$ and {\em inter}-bilayer $(J')$ exchange interactions at each $x$.  At
$x=0.30$, $J_{\parallel}=-4$~meV and $J_{\perp}=-5$~meV, namely almost isotropic and
ferromagnetic.  Upon increasing $x$,
$J_{\perp}$ rapidly approaches zero while $|J_{\parallel}|$ increases slightly,
indicating an enhancement of the planar magnetic anisotropy.  At $x=0.48$,
$J_{\parallel}$ reaches
$-9$~meV, while $J_{\perp}$ turns to $+1$~meV indicating an antiferromagnetic 
interaction.
%
%
Such a drastic change of the exchange interactions can be ascribed to the change of the
relative stability of the $d_{x^{2}-y^{2}}$ and $d_{3z^{2}-r^{2}}$ orbital states upon
doping.  However, a simple linear combination of the two states results in an orbital
state with an orthorhombic symmetry, which is inconsistent with the $I4/mmm$ tetragonal
symmetry of the crystal structure.  We thus propose that an ``orbital liquid'' state 
realizes in LSMO327, where  the charge distribution symmetry is kept tetragonal
around each Mn site.  
%
%
Orbital liquid states are formulated in a theoretical model which takes into account
strong electron correlation.  The calculated results satisfactorily explain the
systematic changes of the  exchange interactions in LSMO327 observed in the experiments. 
\end{abstract}


\pacs{75.30.Ds,75.30.Et,61.12.-q,75.10.-b}


\phantom{.}
]
\narrowtext

\section{Introduction}
\label{Introduction}

%
%

Rare-earth doped Mn perovskite oxide R$_{1-x}$A$_{x}$MnO$_{3}$ (R: rare-earth ion, A:
alkaline-earth ion) is a prototype of the colossal magnetoresistance (CMR) materials. 
Considerable amounts of efforts have been made to clarify the magnetic, electrical
and structural properties of these
systems.\cite{Kusters_89,Chabara_93,Jin_94,Urushibara_95,Hwang_95}  It has become
recognized that the complex and delicate balance among the internal degrees of freedom of
electrons, i.e., charge, spin and orbital, is a key to understand the physics of these
materials.\cite{Tokura_00a,Tokura_00b}  In the Mn perovskite oxides, the orbital degree
of freedom arises from an electron in the doubly degenerated $e_{g}$ states of a
Mn$^{3+}$ ion in a MnO$_{6}$ octahedron.  Comparing with the charge and spin degrees of
freedom, however, the orbital degree of freedom has been much less explored, partially
due to lack of experimental techniques to {\it directly} measure the ordering
processes and the states, with the exception of a pioneering work by Akimitsu and Ito
who have established the orbital ordered state in K$_{2}$CuF$_{4}$ by measuring the
anisotropy of the magnetic form factor using polarized neutrons.\cite{Akimitsu_76}  Recently,
resonant x-ray scattering techniques were successfully applied to the detection of the
orbital ordering process in La$_{0.5}$Sr$_{1.5}$MnO$_{4}$,\cite{Murakami_98a} which was a
significant step toward the understanding of the ``third'' degree of freedom of
electrons.  In the case of LaMnO$_{3}$, the resonant x-ray scattering techniques have
provided direct evidence of the orbital ordering. \cite{Murakami_98b}  Two-dimensional (2D)
planar ferromagnetic coupling found in this three-dimensional (3D) lattice can be naturally
explained by this orbital ordering.\cite{Hirota_96,Moussa_96,Solovyev_96,Ishihara_96} 

%
%

R$_{1-x}$A$_{x}$MnO$_{3}$ consists of MnO$_{6}$ octahedra which are three dimensionally
connected with sharing the corners.  Due to mismatch of ionic radii of Mn and
(R$_{1-x}$A$_{x}$) ions, the Mn-O-Mn bond angle deviates from 180$^{\circ}$, which is
called buckling.\cite{Goodenough_70}  Since doping holes inevitably alter the average
ionic radius of (R$_{1-x}$A$_{x}$) ion, the amount of buckling also changes, resulting in
a variety of crystal structures.\cite{Kawano_96}  Structural phase transitions are also
observed with changing temperature,\cite{Urushibara_95,Caravajal_98} and can be induced
by an external magnetic field.\cite{Asamitsu_95}  The complexity of structural properties
of R$_{1-x}$A$_{x}$MnO$_{3}$ is an interesting issue,\cite{Cox_01} but makes it difficult
to study the role and significance of the orbital degree of freedom in affecting the
magnetic and transport properties, because orbitals are strongly affected by surrounding
structural environment.

%
%

\begin{figure}
\epsfxsize=0.6\columnwidth
\centerline{\epsfbox{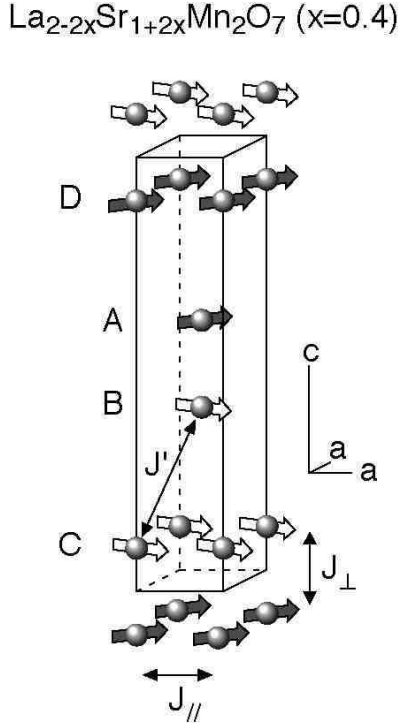}}
\vspace{0.1in}
\caption{Schematic representation of the magnetic spin arrangement on Mn ions in the
$I4/mmm$ tetragonal cell of La$_{1.2}$Sr$_{1.8}$Mn$_{2}$O$_{7}$. Each Mn ions are
surrounded by an O$_{6}$ octahedron.  The lattice parameters are $a=b=3.87$ and
$c=20.1$~\AA~at 10~K.\protect\cite{Hirota_98}.  Notations are explained in
\S.~\ref{Analysis}.}
\label{Fig:Structure}
\end{figure}

\begin{figure}
\epsfxsize=0.9\columnwidth
\centerline{\epsfbox{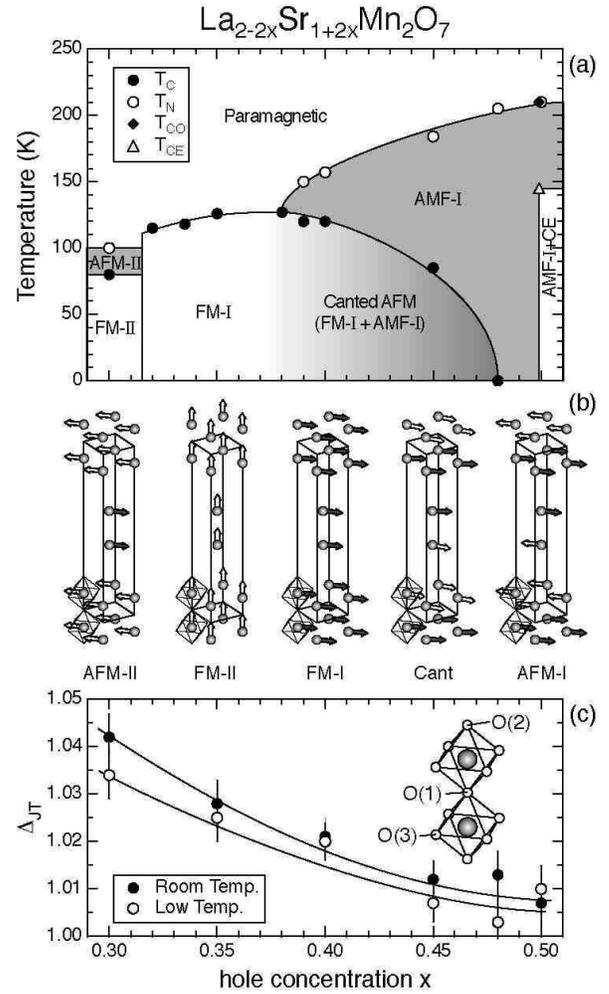}}
\vspace{0.1in}
\caption{(a) Structural and magnetic phase diagram of LSMO327 after Kubota {\it et
al.}\protect\cite{Kubota_00}.  (b) Several different magnetic structures appearing in the
phase diagram are schematically drawn.  (c) Hole concentration dependence of the JT
distortion, which is defined as the ratio of the averaged apical and the equatorial Mn-O
bond lengths, at room temperature and 10~K.}
\label{Fig:Phase_diagram}
\end{figure}

Mn perovskite oxides are generally represented by the Ruddlesden-Popper notation
(R,A)$_{n+1}$Mn$_{n}$O$_{3n+1}$, where the effective dimensionality
can be adjusted by changing the number of MnO$_{2}$ sheets, $n$, blocked with
(R,A)$_{2}$O$_{2}$ layers.  As for the double layered Mn  perovskite $(n=2)$, Moritomo
{\it et al.}\cite{Moritomo_96} have found an extremely large magnetoresistance around
Curie temperature $T_{c}$ in a single crystal of La$_{1.2}$Sr$_{1.8}$Mn$_{2}$O$_{7}$,
which was followed by intensive studies of La$_{2-2x}$Sr$_{1+2x}$Mn$_{2}$O$_{7}$
(LSMO327) and related compounds.\cite{Kimura_00}  Figure~\ref{Fig:Structure} schematically
shows the structure and the magnetically ordered state at $x=0.40$.\cite{Hirota_98}  A
comprehensive magnetic and structural phase diagram of LSMO327 has been established by
Kubota {\it et al.}\cite{Kubota_99a,Kubota_00} in a wide range of $x$ $(0.30 \le x \le
0.50)$ as shown in  Fig.~\ref{Fig:Phase_diagram}(a) through systematic powder
neutron-diffraction studies combined with the Rietveld analysis.  They have found that
LSMO327 exhibits a planar ferromagnetic (FM) structure FM-I in the range $0.32 \le x \le
0.38$ at low temperatures and that a finite canting angle between neighboring layers
starts appearing around $x \sim 0.39$ and reaches 180$^{\circ}$ (AFM-I, i.e., A-type AFM)
for $x \ge 0.48$.  They also found that the magnetic moments are aligned parallel to the
$c$-axis at $x=0.30$, indicating a phase boundary between $x=0.30$ and 0.32.  At $x=0.50$,
the magnetic structure exhibits complicated temperature dependence due to charge
ordering.\cite{Kubota_99b,Argyriou_00}

%
%

In contrast to the rich magnetic phase diagram, the structure of LSMO327 is fairly
simple.  Although anomalous structural behaviors were reported around $T_{c}$
suggesting a strong coupling among charge, spin and
lattice,\cite{Mitchell_97,Argyriou_99} there is only a single tetragonal ($I4/mmm$) phase
in the entire hole concentration ($0.30 \le x \le 1.0$) and temperature ranges ($T \le
400$~K)  studied so far, except a recently discovered orthorhombic 
($Immm$) phase which
 exists in a limited concentration range ($0.75 < x < 0.95$).\cite{Ling_00}   This is
most likely due to the layered structure which absorbs the changes of the Mn-O bond
lengths and the average ionic radius of La and Sr ions upon doping.
%
%
It was also found that the Jahn-Teller (JT) type lattice distortion
$\Delta_{JT}$ of Mn-O$_{6}$ octahedra monotonically changes with increasing
$x$ as shown in Fig.~\ref{Fig:Phase_diagram}(c).\cite{Kubota_00}  Note that $\Delta_{JT}$
is defined by the ratio of the averaged apical Mn-O bond length to the equatorial Mn-O
bond length, i.e., $\Delta_{JT} \equiv ( d_{{\rm Mn-O(1)}}+d_{\rm Mn-O(2)} )/ 2 d_{\rm
Mn-O(3)}$, where $d_{\rm Mn-O}$ is a distance between nearest neighbor (NN) Mn and O
ions.  The positions of the O ions are depicted in Fig.~\ref{Fig:Phase_diagram}(c). The
results are in good agreement with x-ray diffraction
measurements.\cite{Moritomo_98,Okuda_99}  The JT distortion in LSMO327 stabilizes either
$d_{x^2-y^2}$ or $d_{3z^2-r^2}$ state.  The contraction of MnO$_{6}$ octahedron upon
doping implies the stabilization of $d_{x^{2}-y^{2}}$ state, i.e., a pseudo-2D
$e_{g}$ band, in a heavily doped region.
As discussed later, such a structural change itself cannot account for the systematic 
changes of the exchange interactions within the conventional double exchange scenario. 
The dominance of the A-type AFM structure with the decrease of
$\Delta_{JT}$ is ascribed to the change in the $e_{g}$ orbital state from
$d_{3z^{2}-r^{2}}$ to $d_{x^{2}-y^{2}}$.\cite{Hirota_98,Kubota_00}  The importance of the
$e_{g}$ orbital state was also pointed out in striction measurements by Kimura {\it et
al.}\cite{Kimura_98} as well as Argyriou {\it et al.} \cite{Argyriou_99} and Medarde {\it
et al.}\cite{Medarde_99}

%
%

The dynamical magnetic properties of LSMO327 with $x=0.4$ were measured by Fujioka {\it
et al.}\cite{Fujioka_99}  They found that the spin wave dispersion is almost perfectly
2D with the in-plane spin stiffness constant $D \sim 151$~meV\AA.  This 
value is similar to that of  La$_{1-x}$Sr$_{x}$MnO$_{3}$ (LSMO113) with $x \sim 0.3$, 
though $T_{C}$ is three times higher.   They found that there exist two branches due to
a coupling between layers {\em within} a double-layer.   They have analyzed the
spin-wave dispersion and the differential scattering cross section by applying the
Holstein-Primakoff transformation to a Heisenberg Hamiltonian with in-plane
$(J_{\parallel})$ and {\em intra}-bilayer $(J_{\perp})$ interactions (See
Fig.~\ref{Fig:Structure}).  They have estimated that the {\em intra}-bilayer coupling is
$\sim 30$~\% of the in-plane coupling, which is contrary to the fact that the Mn-O bond
lengths are similar. They speculated that $d_{x^{2}-y^{2}}$ orbital is dominant at
$x=0.40$, which enhances the double-exchange, i.e., ferromagnetic, interaction within
the planes.  This interpretation is consistent with the conclusion
drawn by previous structural studies\cite{Kubota_99a,Kubota_00}  The $l$-dependence of the
scattering intensity due to the spin wave show an excellent agreement with the theoretical
calculation of the differential scattering cross section.  Similar inelastic
neutron-scattering experiments were independently performed by Chatterji {\it et
al.},\cite{Chatterji_99a,Chatterji_99b} which gives consistent results with that of
Fujioka {\it et al.}\cite{Fujioka_99} 

%
%

The importance of the $e_{g}$ orbital state in determining the magnetic and transport
properties of LSMO327 is clear.  Moreover, its simple structure makes LSMO327
more favorable platform to study the roles of orbital degrees of freedom than LSMO113. 
The resonant x-ray scattering technique, however, is not directly applicable to LSMO327.
%
%
This is because the orbital state is presumably not antiferro-type long-range ordering as
seen in LaMnO$_{3}$, where the two types of the orbital are alternately aligned, thus the
superlattice reflections due to the orbital ordering do not appear in the resonant x-ray
scattering.  Instead, it is necessary to determine the $e_{g}$ orbital polarization.  In
the present study, we have carried out a series of inelastic neutron-scattering
measurements on single crystals of the layered perovskite
La$_{2-2x}$Sr$_{1+2x}$Mn$_{2}$O$_{7}$ (LSMO327) at $x=0.30, 0.35, 0.40,$ and $0.48$.  To
quantitatively determine the magnetic interactions in LSMO327, we have calculated the
spin-wave dispersion and the differential scattering cross section numerically by
applying the Holstein-Primakoff transformation\cite{Holstein_40} and the Bogoliubov
transformation to a Heisenberg model with the in-plane $(J_{\parallel})$, {\em
intra}-bilayer $(J_{\perp})$ and {\em inter}-bilayer $(J')$ interactions.   We found that
the exchange interactions systematically changes with changing $x$.  Such an $x$
dependence of the exchange interactions is well explained by an orbital liquid picture. 

%
%

\section{Experimental Procedures}
\label{Experimental}

%
%

LSMO327 powder was prepared by solid-state reaction using prescribed amount of pre-dried
La$_{2}$O$_{3}$ (99.9~\%), Mn$_{3}$O$_{4}$ (99.9~\%) and SrCO$_{3}$ (99.99~\%).  The
powder mixture was calcined in the air for 3 days at 1250 --1400$^{\circ}$C with
frequent grinding.  The calcined powder was then pressed into a rod and heated at
1450$^{\circ}$C for 24~h.  Single crystals were melt-grown in flowing 100~\% O$_{2}$ in
a floating zone optical image furnace with a travelling speed of 15~mm/h.  We powderized
a part of single crystals and performed x-ray diffraction, which shows no indication of
impurities.   Some of the crystals were also checked with electron probe microanalysis 
(EPMA), which indicates no particular spatial inhomogeneity within the instrumental
error.  The $x=0.30$ sample was examined by inductively coupled plasma (ICP) analysis,
which revealed that the ratio of La, Sr and Mn is 28.6 : 32.2: 39.2, which is in good
agreement with the ideal ratio, 28.0: 32.0: 40.0.  These analyses indicate that our
samples are sufficiently stoichiometric and homogeneous.  All the single crystals
studied are domain samples with mosaic spread of 0.3 -- 0.8$^{\circ}$
full-width-at-half-maximum (FWHM).  The samples have a cylindrical shape, which size is
typically $5~\phi \times 20-30$~mm.  Similarly grown samples in the range $0.3 \le x \le
0.5$ were powderized and studied in detail by powder neutron diffraction techniques and
the Rietveld analysis.  The results were already published in
Refs.\onlinecite{Kubota_99a,Kubota_00}.  The transport properties of the samples were also
measured, part of which were published in Refs.\onlinecite{Kubota_00b,Moritomo_00}.  The
results are consistent with previously reported data.\cite{Moritomo_96}

%
%

Neutron scattering measurements were carried out using the Tohoku University triple-axis
spectrometer TOPAN located in the JRR-3M reactor of the Japan Atomic Energy Research
Institute (JAERI).  The spectrometer was set up in the standard triple-axis mode with
the fixed final energy at 14.7~meV and the horizontal collimation of
Blank-60$'$-S-60$'$-Blank. The $(0\ 0\ 2)$ reflection of pyrolytic graphite (PG) was
used to monochromate and analyze the neutron beam, together with a PG filter to
eliminate higher order contamination.  The sample was mounted in an Al can so as to give
the $(h\ 0\ l)$ zone in the tetragonal $I4/mmm$ notation.  The Al can was then attached
to the cold finger of a closed-cycle He gas refrigerator.  All the data were taken at
10~K.

%
%

\section{Results}
\label{Results}

%
%

As shown by Fujioka {\it et al.}\cite{Fujioka_99} and Chatterji {\it et
al.},\cite{Chatterji_99a,Chatterji_99b} the spin wave dispersion of LSMO327 with $x=0.4$
should have two modes, i.e., acoustic (A) and optical (O) branches, due to a coupling
between layers within a double-layer.  It was theoretically shown that the A-branch has
maximum intensity at $l=5n$ ($n$: integer), while the phase of the O-branch is shifted by
$\pi$ in the double layered system.\cite{Fujioka_99}  We thus measured the spin-wave
dispersions along $[h\ 0\ 0]$ around $(1\ 0\ 0)$ and $(1\ 0\ 5)$ for the A-branch and
around $(1\ 0\ 2.5)$ and $(1\ 0\ 7.5)$ for the O-branch.  To study the differential cross
sections of spin waves, we have also measured the
$l$-dependence of the spin-wave intensities of A and O branches at a fixed transfer
energy $\Delta E=E_{i}-E_{f}$.

%
%

\begin{figure}
\epsfxsize=0.75\columnwidth
\centerline{\epsffile{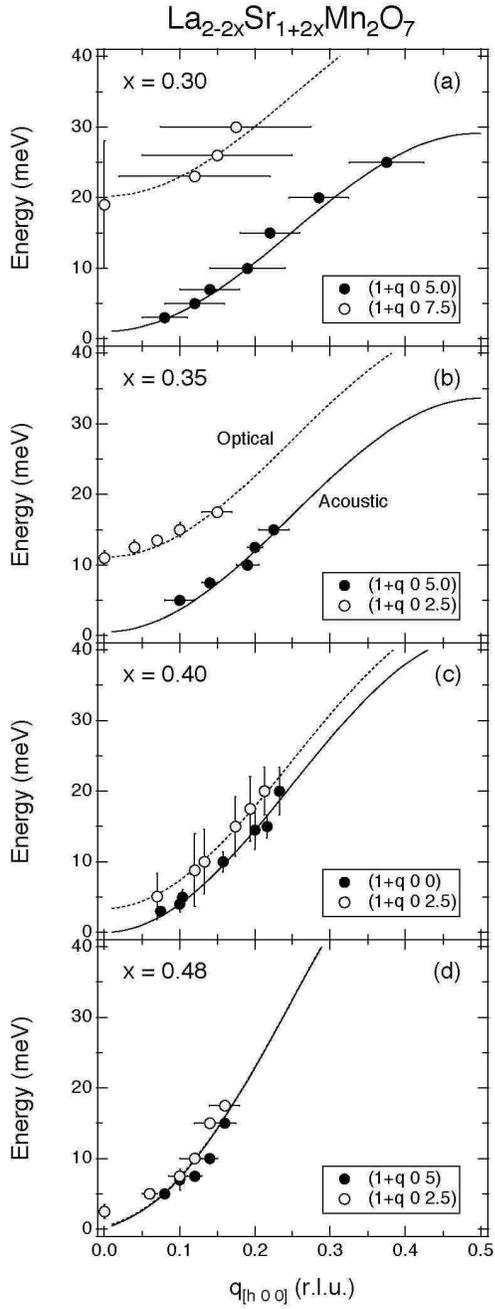}}
\vspace{0.1in}
\caption{The dispersion relations of spin waves at 10~K for (a) $x=0.30$, (b) $x=0.35$,
(c) $x=0.40$, and (d) $x=0.48$.  Error bars correspond to the FWHM of peak profiles. 
Solid circles and open circles indicate the acoustic branch and
the optical branch, respectively.  Solid and dotted curves are obtained by fitting to
theoretical models described in \S.~\ref{Analysis}.}
\label{Fig:Dispersion}
\end{figure}

\begin{figure}
\epsfxsize=0.75\columnwidth
\centerline{\epsffile{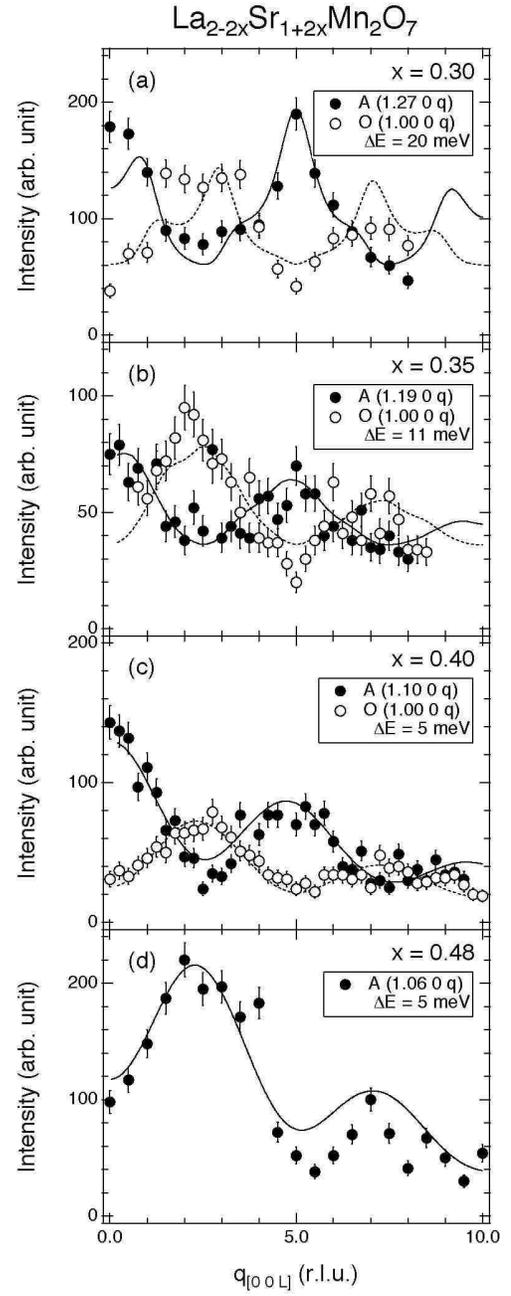}}
\vspace{0.1in}
\caption{Differential cross sections along the out-of-plane direction, which were
obtained from the $l$-dependence of the constant-$E$ scans.  The solid and open circles
indicate intensities of the acoustic and optical branches, respectively.  Solid and dotted
curves are obtained by fitting to
theoretical models described in \S.~\ref{Analysis}.  The acoustic branch is dominant at
(1~0~0) and (1~0~5), and the optical branch is dominant at (1~0~2.5) and (1~0~7.5).}
\label{Fig:Cross_section}
\end{figure}

Figure~\ref{Fig:Dispersion}(a) shows the dispersion relations of spin waves at 10~K for
$x=0.30$.  Error bars correspond to the full-width at half-maximum (FWHM) of peak
profiles including the instrumental resolution.  Spin waves of the A-branch are well
defined in the low $q$ and low energy region.  However, the O-branch exhibits a large
broadening even at the magnetic zone center.  The $l$-dependence of the constant energy
scan at $\Delta E=20$~meV are shown in Fig.~\ref{Fig:Cross_section}(a).  As expected, the
A and O branches exhibit intensity maxima at $l=5n$ and $l=\frac{5}{2} (2n+1)$.  Solid and
dotted curves are fitting to theoretical calculations, which is described in the next
section.

%
%

Figure~\ref{Fig:Dispersion}(b) shows the dispersion relations for $x=0.35$.  The
dispersion curves of both the A and O branches become slightly steeper than those of
$x=0.30$, indicating that the in-plane magnetic interaction $J_{\parallel}$ increases
only gradually.  However, the gap between the A and O branches becomes almost half of
that at $x=0.30$.  Since the gap corresponds to the out-of-plane magnetic interaction
$J_{\perp}$, this result indicates that $J_{\perp}$ decreases considerably.  More
quantitative analysis will be made in the following sections.  We have also noticed that
spin waves are fairly well defined below 20~meV, and that they becomes significantly
broad above 20~meV, even in the constant energy scans.  The $l$-dependence of the
constant energy scan at $\Delta E=11$~meV are shown in Fig.~\ref{Fig:Cross_section}(b).

%
%

Figure~\ref{Fig:Dispersion}(c) shows the dispersion relations for $x=0.40$. Part of the
data has been already reported.\cite{Fujioka_99}  The $l$-dependence of the constant
energy scan at $\Delta E=5$~meV are shown in Fig.~\ref{Fig:Cross_section}(c).  Following
the tendency between $x=0.30$ and 0.35, the dispersion curves become steeper and the gap
becomes smaller.   As shown in Fig.~\ref{Fig:Dispersion}(c), the energy-width shows
anomalous broadening near the zone boundary, which was found by Fujioka {\it et
al.}\cite{Fujioka_99}  Furukawa and Hirota\cite{Furukawa_00} investigated this broadening
from both theoretical and experimental point of view, and ascribed it to a strong
magnon-phonon coupling.  Let us consider a dispersionless optical phonon branch at
$\hbar\Omega_{0}$, and a spin wave dispersion $\hbar\omega(q)$.  When a magnon with
momentum $q$ has energy $\omega(q) > \Omega_{0}$, it is possible to find an inelastic
channel to decay into a magnon-phonon pair with momentum $q'$ and $q-q'$, respectively,
which satisfies the energy conservation law, $\omega(q)=\omega(q')+\Omega_{0}$.  This
decay channel gives rise to an abrupt broadening of the line-width of the spin wave
branch which crosses the optical phonon.

%
%

Figure~\ref{Fig:Dispersion}(d) shows the dispersion relations for
$x=0.48$.  The $l$-dependence of the constant energy scan at $\Delta E=5$~meV are shown
in Fig.~\ref{Fig:Cross_section}(d).  

%
%
The two branches measured at (1\ 0\ 5) and (1\ 0\ 2.5) nearly degenerate. 
Unlike $x=0.30$, 0.35 and 0.40, $x=0.48$ has the A-type AFM (AFM-I) structure, resulting
in a fundamental difference in the spin wave dispersion.  We will discuss this difference
in detail in the next section.
%
%
%

%
%

As pointed out by Furukawa and Hirota, there exists an optical phonon branch around
$\Delta E=20$~meV for $x=0.40$, which we have also confirmed for the other compositions we
studied in the present work.  We have noticed that the line-width of spin waves above
this characteristic energy of 20~meV become significantly broaden, which is consistent
with the strong magnon-phonon coupling model mentioned above.  Khaliullin and
Kilian\cite{Khaliullin_00} considered an orbitally degenerate double-exchange system
coupled to Jahn-Teller active phonons, which explains the softening of spin waves at the
zone boundary found in various ferromagnetic manganese oxides.  Their model could be
applicable to the anomalous broadening of the spin waves of LSMO327 near the zone
boundary.  However, the large line-width of the O branch near the {\em zone center} may
not be accounted for because their theoretical model does not affect the small-momentum
spin dynamics.  We thus believe that there exists a significantly strong magnon-phonon
coupling as suggested.    In the present paper, we have combined constant $q$ and energy
scans to efficiently measure the dispersion relations, which is our principle target of
the present work.  Constant energy scans are particularly useful to avoid contamination
from dispersionless optical phonon branches.  To further investigate this issue, however,
it is necessary to measure the energy widths at various $q$ utilizing constant
$q$ scans, which we plan to carry out in the next step.

\section{Analysis}
\label{Analysis}

In order to analyze the experimental results of the  spin-wave dispersion relation  and
the scattering cross section in LSMO327,  we start from the Heisenberg model where $3d$
electrons in a Mn ion are  treated as localized spins.  In the FM (A-type AFM) structure
for LSMO327 with $x=$0.3, 0.35 and 0.4  ($x=$0.48),   a magnetic unit cell includes two
(four) Mn ions termed $A$ and $B$ ($A$, $B$, $C$, and $D$).  Between Mn ions, three kinds
of the exchange interactions, i.e.  the in-plane ($J_{\parallel}$),  intra-bilayer
($J_{\perp}$) and  inter-bilayer ($J'$) exchange interactions are introduced.  The
schematic picture is shown in Fig.~\ref{Fig:Structure}.  The Hamiltonian is given by 
\begin{eqnarray}
{\cal H} & = & {1 \over 2} \sum_{i l} \vec S^{l}(\vec r_{il}) \biggl \{ 
J_{\parallel} \sum_{\delta_{\parallel}}  \vec S^l           (\vec r_{il}+\vec \delta_{\parallel})
\nonumber \\ &   & +
J_{\perp}     \sum_{\delta_{\perp}}      \vec S^{l_{\perp}} (\vec r_{il}+\vec \delta_{\perp})+
J'            \sum_{\delta'}             \vec S^{l'}        (\vec r_{il}+\vec \delta')
\biggr \} , 
\label{eq:hamiltonian}
\end{eqnarray}
where $\vec S^l(\vec r_{il})$ is the spin operator at a Mn ion $l$ in the $i$-th unit
cell  and $\vec r_{il}$ is a position of the ion.   Spin quantum number is assumed to be
$S=2(1-x)+{3 \over 2}x$ with  a hole concentration $x$.  
$l_{\perp}=l'=(B,A)$ for $l=(A,B)$ in the FM structure  and $l_{\perp}=(B,D,A,C)$ and
$l'=(D,C,B,A)$ for $l=(A,B,C,D)$ in the A-type AFM one. 
$\vec \delta_{\parallel}$, $\vec \delta_{\perp}$ and $\vec \delta'$ indicate  the vectors
connecting the nearest neighboring (NN) Mn ions; 
$\vec \delta_{\parallel}=(\pm a,0,0)$ and $(0, \pm a,0)$ where $a$ is a distance between
NN Mn ions in the $ab$ plane. 
$\vec \delta_{\perp}=(0,0, c_{\perp})$ and $\vec \delta'=(\pm a/2,\pm a/2,-c')$ for $l=B$
and $D$, and 
$\vec \delta_{\perp}=(0,0,-c_{\perp})$ and $\vec \delta'=(\pm a/2,\pm a/2, c')$ for $l=A$
and $C$,   where $c_{\perp}$ and $c'$ are distances between NN MnO$_2$ layers and NN
bilayers, respectively. 
By applying the Holstein-Primakoff transformation to Eq.~(\ref{eq:hamiltonian}),  the
Hamiltonian is rewritten as follows, 
\begin{equation}
{\cal H}=\sum_k \psi^\dagger (\vec k) \varepsilon (\vec k) \psi(\vec k) . 
\label{eq:dis}
\end{equation}
Here, $\psi(\vec k)=(a_k, b_k)$ for the FM structure and  $\psi(\vec k)=(a_k,
b_k^\dagger, c_k , d_k^\dagger)$  for the A-type AFM structure.
$a_k$, $b_k$, $c_k$ and $d_k$ are the boson operators for  the spin operators $\vec S^A$,
$\vec S^B$, $\vec S^C$ and $\vec S^D$, respectively.
$\varepsilon (\vec k)$ is given by 
\begin{equation}
\varepsilon(\vec k)= \pmatrix{
{\rm x}      & {\rm y} \cr
{\rm y}^\ast & {\rm x} \cr
} , 
\end{equation}
with  
\begin{eqnarray}
{\rm x} & = & -4J_{\parallel}S \biggl \{1-{1 \over 2}(\cos ak_{\rm x}+\cos ak_{\rm y})
\biggr\} \nonumber \\ & & 
-J_{\perp}S-4J'S , 
\end{eqnarray}
and
\begin{equation} 
{\rm y}=J_{\perp}Se^{i k_{\rm z} c_{\perp}}+4J' S\cos \biggl( {ak_{\rm x} \over 2} \biggr)
\cos \biggl( {ak_{\rm y} \over 2} \biggr) e^{i k_{\rm z} c'} , 
\end{equation}
for the FM structure, 
and 
\begin{equation}
\varepsilon(\vec k)= \pmatrix{
X   &   Y^\ast&   Z &        \cr
Y   &    X    &     & Z^\ast \cr
Z^\ast& & X & Y \cr
& Z & Y^\ast & X \cr
} , 
\end{equation}
with  
\begin{eqnarray}
X & = &-4J_{\parallel} S \biggl \{ 1-{1 \over 2}(\cos ak_{\rm x}+\cos ak_{\rm y}) \biggr \}
\nonumber \\ & &
+J_{\perp}S-4J'S , 
\end{eqnarray}
\begin{equation} 
Y=J_{\perp}Se^{i k_{\rm z} c_{\perp}} , 
\end{equation}
and 
\begin{equation} 
Z=4J'S \cos \biggl( {ak_{\rm x} \over 2} \biggr)
       \cos \biggl( {ak_{\rm y} \over 2} \biggr) e^{i k_{\rm z} c'} , 
\end{equation}
for the A-type AFM structure.   By utilizing the canonical transformation of the
Hamiltonian Eq.~(\ref{eq:dis}),  the dispersion relations of the spin waves are
obtained.   As for the FM structure, in particular, the dispersion relations are  given
analytically as $\omega_k={\rm x}\pm | {\rm y} |$.  On an equal footing, the differential
scattering cross section for the inelastic-neutron scattering from spin wave is given by  
\begin{eqnarray}
\lefteqn{
{d^2 \sigma \over d \Omega d \omega'} =
 {\gamma e^2 \over m c^2}   
\biggl ( {1 \over 2} g F(\vec Q)     \biggr )^2
{k' \over k} e^{-2W(\vec Q)} N {S \over 8}
}  &  &\nonumber \\
& \times & 
\sum_{ll'm} \sum_{q G} \biggl \{
\delta (\omega-\omega_q^m) \delta(\vec Q-\vec G-\vec q) (1+n_q^m) 
U_{lm}^\dagger (\vec q) U_{ml'} (\vec q)  \nonumber \\
& + &
\delta (\omega+\omega_q^m) \delta(\vec Q-\vec G+\vec q) n_q^m 
U_{lm} (\vec q) U_{ml'}^\dagger (\vec q)  \biggr \} , 
\label{eq:cs}
\end{eqnarray}
where $\vec q$ and $\omega_q^m$ are the momentum and energy of spin wave of the mode $m$,
respectively,  and $n_q^m=1/(e^{\beta \omega_q^m}-1)$ is a Bose factor with temperature
$T=1/\beta$.  $U(\vec q)$ is a matrix introduced in the canonical transformation, $\vec
Q=\vec k_{i}-\vec k_{f}$ is the momentum transfer with $\vec k_{i}$ ($\vec k_{f}$) being
the momentum of the incident (scattered) neutron, and $\vec G$ is a reciprocal lattice
vector.  $F(Q)$ and $W(Q)$ are the magnetic structure factor and the Debye-Waller factor,
respectively.  

The experimental results of the dispersion relation and the  differential scattering
cross section in LSMO327 are  fitted by utilizing the least-squares method.   The
calculated results are shown in Figs.~\ref{Fig:Dispersion} and \ref{Fig:Cross_section}
together with the experimental data.  Note that  the dispersion relation along
$[h\ 0\ 0]$ and  the cross section along $[0\ 0\ l]$ are  sensitive to $J_{\parallel}$
and $J_{\perp}$, and $J_{\perp}$ and $J'$,  respectively.   In $x=0.3$, 0.35 and 0.4, the
A and O branches are well separated.  These two correspond to the in-phase and out-of
phase motions of spins in  NN MnO$_2$ layers.   The energy separation between the two
branches at the point $\Gamma$ and the stiffness constant of the A-branch in the $ab$
plane are given  by $-2S(J_{\perp}+4J')$ and
$D=-SJ_{\parallel}$, respectively.   An intensity oscillation along $[0\ 0\ l]$ is
factorized by the functions $1+\cos(c_{\perp}q_z)$ and $1-\cos(c_{\perp}q_z)$ for the A
and O branches, respectively, where $c_{\perp} \sim c/5$ with $c$ being the lattice
constant in the $c$ axis.  This is attributed to the spin correlation between NN MnO$_2$
layers controlled by $J_{\perp}$.   Additional fine structures in the intensity are
caused by $J'$.  On the contrary,  in the A-type AFM structure,  four modes of the spin
wave exist  and separate into the A and O branches corresponding to  the in-phase and
out-of phase motions of spins between NN bilayers.  Here,  the each branch is doubly
degenerate and the energy separation between the two is of the order of $J'$.  Since $J'
\sim J_{\parallel}/1000$ at $x=0.48$, as mentioned later,   these cannot be observed
separately by the experiments.  An intensity oscillation along [0\ 0\ l]   is factorized
by a function $1-\cos(c_{\perp}q_z)$ originating from  the antiferromagnetic spin
alignment between the NN MnO$_2$ layers.

The $x$ dependence of the exchange interactions is  shown in Fig.~\ref{Fig:Exchange}.  
All interactions systematically change with $x$;  with increasing $x$ from 0.3,
$|J_{\parallel}|$ increases, 
$J_{\perp}$ rapidly approaches to zero and  changes its sign from negative to positive. 
$|J'|$ decreases with $x$, although its value is 10$\sim$1000 times smaller than
$|J_{\parallel}|$ and $|J_{\perp}|$.   The systematic change of the interactions 
correlates with that of the lattice distortion  in a MnO$_6$ octahedron represented by
$\Delta_{JT}$ (Fig.~\ref{Fig:Phase_diagram}(c)).  However, this is an opposite direction
predicted by  the conventional double-exchange scenario,  where $|J_{\parallel}|$ is
reduced with increasing $x$  because magnitude of the double-exchange interaction is
proportional to  the hopping integral between Mn ions in the strong Hund-coupling
limit.   Therefore, a structural change itself   cannot account for that of the exchange
interactions.  Let us taking into account the orbital degree of freedom in a Mn ion.   In
LSMO327 with hole concentration $x$, $1-x$ electrons occupy the two $e_g$ orbitals. 
Character of the occupied orbital controls the anisotropy of the hopping integral of
electrons, i.e.,  that of the ferromagnetic double exchange interaction.   The systematic
change of the exchange interactions can be  explained by assuming that the $3d_{x^2-y^2}$
orbital is relatively stabilized with increasing $x$.  The exchange interaction between
Mn ions  is sum of the ferromagnetic double-exchange interaction and the 
antiferromagnetic superexchange one $J_{AFM}$ acting between $t_{2g}$ spins.  The more the
$d_{x^2-y^2}$ orbital is stabilized, the more the ferromagnetic interaction  in $ab$
plane (along the
$c$ axis) becomes strong (weak).  Then $J_{AFM}$ overcomes the ferromagnetic interaction
along the $c$ axis as shown in the region of $x=0.4-0.48$.  This assumption for the
orbital stability is consistent with the $x$ dependent $\Delta_{JT}$
\cite{Kubota_00,Kimura_98,Medarde_99} and is supported by the previous theoretical work
where the stability of the orbitals  is examined by the Madelung potential calculation.
\cite{Akimoto_99,Okamoto_01}

\section{Summary and Discussion}

We discuss possible orbital states in LSMO327 with $0.3 \le x<0.5$ and its relation to
the anisotropy of the ferromagnetic interaction in more detail.  The orbital state at
each Mn ion is represented by the pseudo-spin operator defined by 
\begin{equation} 
T_{i \mu} = {1 \over 2} \sum_{s \gamma \gamma'} 
d_{i \gamma s}^\dagger \sigma_\mu d_{i \gamma' s} , 
\label{eq:ps}
\end{equation}
for $\mu=(x,z)$.  $d_{i \gamma s}$ is the annihilation operator for the $e_g$ electron at
site $i$ with spin $s$ and orbital $\gamma$, and $\sigma_\mu$ are the Pauli matrices.  
In the eigen state of $T_{i z}=+(-)1/2$,   an electron occupies the $d_{3z^2-r^2}$
($d_{x^2-y^2}$) orbital at site $i$.  $T_{iz}$ ($T_{ix}$)  describes the charge
quadrupole moment with tetragonal (orthorhombic) symmetries and couples  with the lattice
distortion with the same symmetry;  
\begin{equation}
{\cal H}_{JT}=-g\sum_{i \mu=x,z}T_{i \mu }Q_{i \mu} , 
\end{equation}
where $Q_{i z}$ ($Q_{i x}$) describes the O ion distortions in a MnO$_6$ octahedron.  The
orbital ordered state is characterized by a magnitude and an angle of this operator,
i.e.  $|\langle \vec T \rangle|=\sqrt{\langle T_x \rangle^2+\langle T_z \rangle^2}$  and
$\Theta=\tan^{-1}(\langle T_x \rangle/\langle T_z \rangle)$   where $\langle \cdots
\rangle$ is the thermal average.  For example, in the $(d_{3x^2-r^2},d_{3y^2-r^2})$-type
orbital ordered state  observed in LaMnO$_3$, 
$\Theta=2\pi/3$ and $-2\pi/3$ for the Mn sites where the $d_{3x^2-r^2}$ and
$d_{3y^2-r^2}$  orbitals are occupied, respectively. 
\par
As mentioned in the previous section,  a relative weight of the occupied $d_{x^2-y^2}$
orbital increases continuously  with increasing $x$ from 0.3 to 0.48 in LSMO327  where
the crystal structure remains to be tetragonal ($I4/mmm$).  That is, $\langle T_z (\vec
k=0) \rangle$  is gradually reduced with keeping the condition $\langle T_x (\vec
k=0)\rangle=0$  where $\vec T(\vec k)=1/N \sum_i e^{i \vec k \cdot \vec r_i} \vec T_i$
with the number of the Mn ion $N$ and the position of the $i$-th ion $\vec r_i$.  This
cannot be satisfied by the uniform orbital ordered state where one kind of orbital 
characterized by $\Theta$ is occupied at all Mn sites.  This is because the change of the
orbital state is represented by  the rotation of $\langle \vec T \rangle$  in the
$\langle T_z \rangle$-$\langle T_x \rangle$ plane.   One may think that the
antiferro-type orbital ordered state explains the experiments, when the condition
$\Theta_A=-\Theta_B$, with $\Theta_{A(B)}$ being the angle  in the orbital space for the
$A$ $(B)$ sublattice, is satisfied.
However, this is ruled out by the experimental fact
that the expected superlattice reflection was not reported in $0.3 \le x < 0.5$ by the x-ray  and
electron diffractions.\cite{Li_99,Wakabayashi_00}

\begin{figure}
\epsfxsize=0.75\columnwidth
\centerline{\epsffile{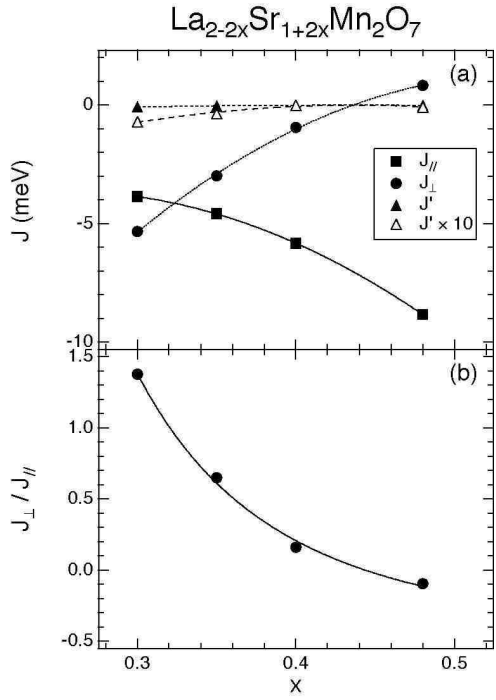}}
\vspace{0.1in}
\caption{(a) $x$ dependence of the exchange interactions obtained by analyses of the
dispersion relations and  the scattering cross sections in the inelastic neutron scattering
experiments. (b) $x$ dependence of the ratio of the exchange interactions $J_{\perp}$ and
$J_{\parallel}$.}
\label{Fig:Exchange}
\end{figure}

\begin{figure}
\epsfxsize=0.75\columnwidth
\centerline{\epsffile{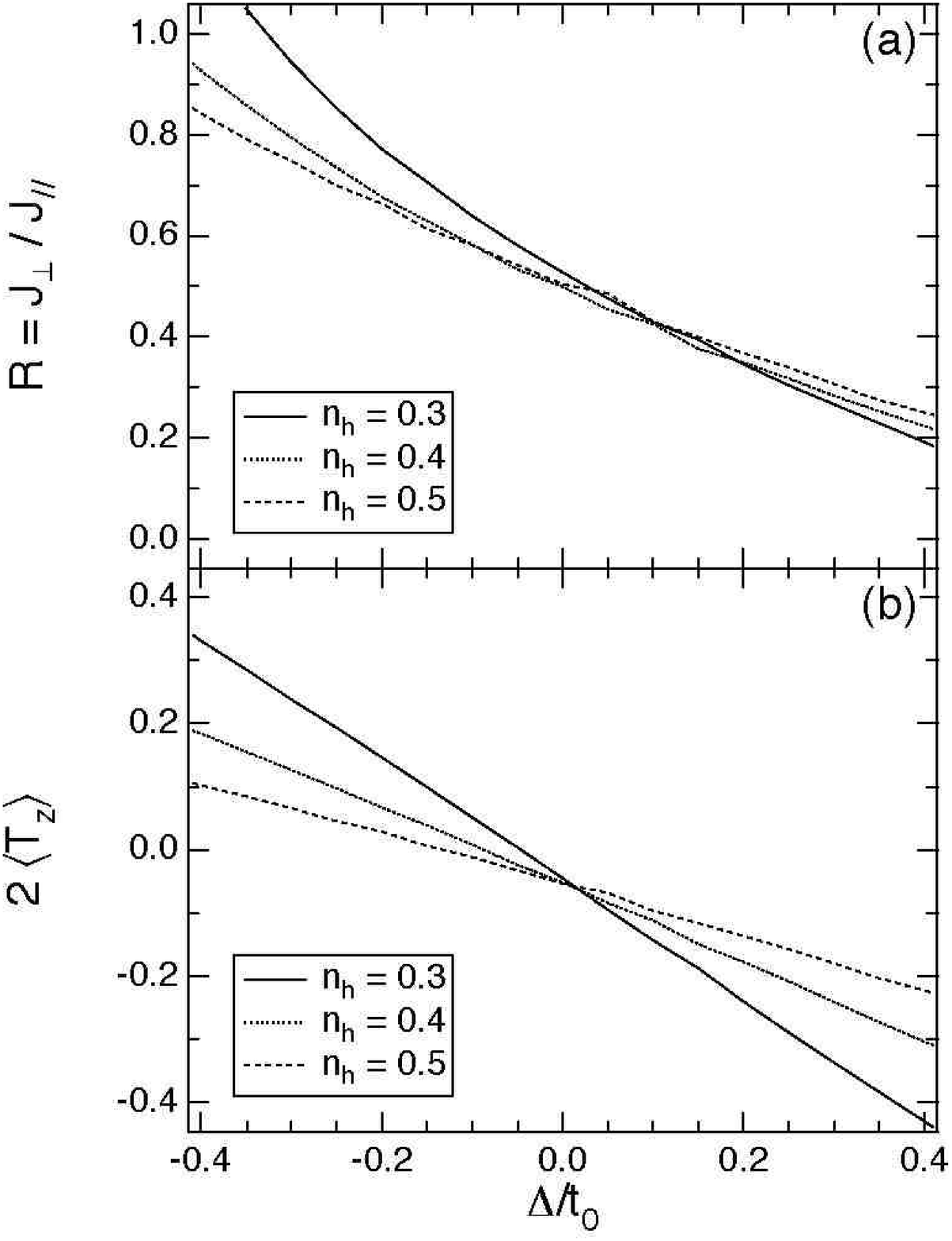}}
\vspace{0.1in}
\caption{
(a) Theoretical results of the exchange interactions ratio R=$J_{\perp}/J_{\parallel}$.  
$J_{\parallel}$ and $J_{\perp}$ are the exchange interactions between NN Mn sites in
the $ab$ plane and along the $c$ direction, respectively.
(b) The relative number of occupied electrons in two $e_g$ orbitals 2$\langle T_z \rangle$. 
}
\label{Fig:Ratio}
\end{figure}

\begin{figure}
\epsfxsize=0.4\columnwidth
\centerline{\epsffile{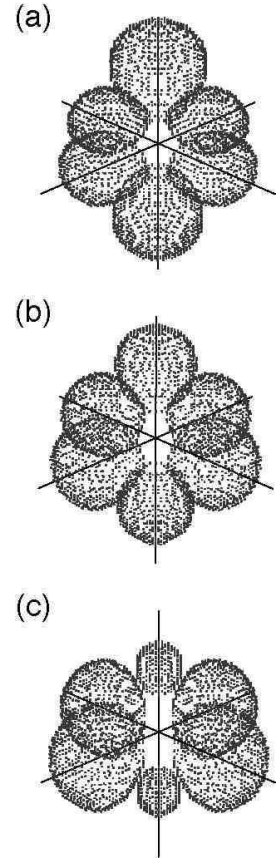}}
\vspace{0.1in}
\caption{Schematic pictures of the charge distribution for $e_g$ electrons in the case of
(a) $\Delta=-0.4$,  (b) 0 and (c) 0.4.}
\label{Fig:Orbital}
\end{figure}

One of the possible orbital states realized in LSMO327 is an {\em orbital liquid} state. 
This state was originally proposed in the ferromagnetic metallic state in LSMO113 by
Ref.~\onlinecite{Ishihara_97}, where  the orderings of both $T_z $ and $T_x $ are
suppressed by the low dimensional character of the  orbital fluctuation.  In the case of
LSMO327, $\langle T_z \rangle$ is finite due to the layered crystal structure.  On the
other hand, $T_{ix}$  does not show ordering and symmetry of the charge distribution
remains to be tetragonal at each Mn site.   In order to formulate this orbital state, 
let us start from the Hamiltonian where the intra-site Coulomb interactions in Mn ions are 
taken into account; 
\begin{eqnarray}
{\cal H} & = & \sum_{\langle ij \rangle \sigma} (t_{ij}^{ \gamma \gamma'} 
\widetilde d_{i \gamma \sigma}^\dagger \widetilde d_{j \gamma' \sigma} + H.c.) 
\nonumber \\
& & - J_H \sum_i \vec S_i \cdot \vec S_{t i}+\Delta \sum_i T_{i z} , 
\label{eq:ham}
\end{eqnarray}
where $\widetilde d_{i\gamma \sigma}=d_{i \gamma \sigma} (1-n_{i \gamma {\bar
\sigma}})(1-n_{i {\bar
\gamma} \sigma}) (1-n_{i {\bar \gamma} {\bar \sigma}})$  is the annihilation operator of
an $e_g$ electron  excluding the doubly occupied states of electrons due to the strong
Coulomb interaction.  
$\vec S_i$ is the spin operator for an $e_g$ electron  defined by $\vec S_i={1 \over 2}
\sum_{s s'
\gamma} d_{i \gamma s}^\dagger \vec \sigma_{s s'} d_{i \gamma s'}$ and $\vec S_{t i}$ is
the spin operator for
$t_{2g}$ electrons  with $S=3/2$. The first and second terms in Eq.~(\ref{eq:ham})
represent  the hopping of $e_g$ electrons between NN Mn sites  and the Hund coupling
between $e_g$ and
$t_{2g}$ spins, respectively.  In the third term, $\Delta$ describes the splitting of the energy
levels of
$d_{3z^2-r^2}$  and $d_{x^2-y^2}$ orbitals due to the tetragonal distortion $\Delta_{JT}$ 
of a MnO$_6$ octahedron.  It is shown from the theoretical calculation in 
Ref.~\onlinecite{Okamoto_01} that $\Delta$ 
monotonically decreases with increasing $x$ for
LSMO327  and its maximum value is of the order of 0.5 eV.  Instead of the actual crystal
structure of LSMO327, a pair of the 2D sheets where a squared lattice
consists of the Mn ions is introduced  because of the weak inter-bilayer exchange
interaction.  
We adopt the slave-boson scheme where 
$\widetilde d_{i\gamma \sigma}$ is decomposed into a product of operators: 
$\widetilde d_{i\gamma \sigma}=f_i^\dagger \tau_{i \gamma} s_{i \sigma}$  where $f_i$ and
$s_{i \sigma}$ are bosonic operators for charge and spin degrees of freedom, 
respectively, and $\tau_{i \gamma}$ is a fermionic one for orbital  associated with the
constraint of 
$\sum_\sigma s_{i \sigma}^\dagger s_{i \sigma}=\sum_\gamma \tau_{i \gamma}^\dagger
\tau_{i \gamma}$  and 
$f_i^\dagger f_i+\sum_\sigma s_{i \sigma}^\dagger s_{i \sigma}=1$ at each site.  The mean
field approximation is introduced; $\langle f_i^\dagger f_j \rangle=x$  and
$\sum_{\sigma} \langle s_{i \sigma}^\dagger s_{j \sigma} \rangle=(1-x)\varepsilon_{ij}$
with  
$\varepsilon_{ij}=+(-)1$ for a ferromagnetic (antiferromagnetic) bond in the orbital part
of the   mean-field Hamiltonian.  It is well known that the slave-boson mean-field
approximation is suitable  to describe the spin liquid state.\cite{Tsvelik_95}  The ratio
of the ferromagnetic exchange interaction in the $ab$ plane to that in the $c$ axis  is
given by $R=J_{\parallel}/J_{\perp}=\chi_{ab}/\chi_c$ with   
$\chi_{l}=\sum_{\gamma \gamma'} \langle \tau_{i \gamma}^\dagger \tau_{i+l \gamma'}
\rangle$ for $l=ab$ and $c$.   The results are shown in Fig.~\ref{Fig:Ratio}(a) where 
$J_H$ is assumed to be infinite and $t_0$ is the hopping integral between 
$d_{3z^2-r^2}$ orbitals in the $c$ axis. 
$R$ continuously decreases with increasing $\Delta$ implying that $d_{x^2-y^2}$ orbital
becomes stable relatively.  This feature does not depend on the hole concentration $n_h$
in the calculation.  Since $\Delta$ is expected to continuously decreases with $x$ in LSMO327, 
the results in Fig.~\ref{Fig:Ratio}(a) explain the experimental results presented in
Fig.~\ref{Fig:Exchange}(b). 
$\Delta$ dependence of $\langle T_x \rangle$  are shown in Fig.~\ref{Fig:Ratio}(b).  
We note that $\langle T_z \rangle$ is zero. It is clearly shown that 
the continuous change of $R$ is controlled by the
character of the occupied orbital.  We also present the schematic pictures of the spatial
distribution of the electronic charge  at a Mn site for the proposed orbital liquid state in
Fig.~\ref{Fig:Orbital}.   The charge distributions have tetragonal symmetry and are not
represented by any linear combination of the atomic wave functions of the $d_{3z^2-r^2}$
and $d_{x^2-y^2}$ orbitals.

To summarize, we have systematically studied low-temperature spin dynamics of the
double-layered perovskite La$_{2-2x}$Sr$_{1+2x}$Mn$_{2}$O$_{7}$ ($0.3 \le x < 0.5$). 
The acoustic and optical branch of the 2D spin-wave dispersion relations
as well as characteristic intensity oscillations along the out-of-plane direction are
successfully explained by theoretical calculations assuming the Heisenberg models with the
in-plane $(J_{\parallel})$, {\em intra}-bilayer
$(J_{\perp})$ and {\em inter}-bilayer $(J')$ exchange interactions.  We have found that
the ratio $R=J_{\parallel}/J_{\perp}$ drastically decreases upon doping holes, which
indicates that the $d_{x^2-y^2}$ orbital becomes more stable than the $d_{3z^2-r^2}$
orbital.   Since a simple linear combination of the two states
results in an orbital state with an orthorhombic symmetry, inconsistent with the
$I4/mmm$ tetragonal symmetry,  we have introduced an ``orbital
liquid'' state, in which the charge distribution symmetry is kept tetragonal around each
Mn site.

\begin{acknowledgments}
Authors would like to thank S.~Okamoto, G.~Khaliullin, A.~Koizumi, Y.~Murakami and
K.~Takahashi for their valuable discussions.   This work was supported by the Grant in Aid
from Ministry of Education,  Science and Culture of Japan, CREST, NEDO, and Science and
Technology Special  Coordination Fund for Promoting Science and Technology.  Part of the
numerical calculation was performed in the HITACS-3800/380  supercomputing facilities in IMR,
Tohoku University. 
\end{acknowledgments}

%

%
%

%
%


\end{document}